\documentclass[sigconf, anonymous=false, screen, review=false, nonacm=true]{acmart}

\usepackage{booktabs}
\usepackage[linesnumbered,ruled,lined,boxed]{algorithm2e}
\usepackage{amsthm}
\usepackage[capitalise]{cleveref}
\usepackage{graphicx}%
\usepackage{placeins} %
\usepackage{color, colortbl}
\usepackage[caption=false, font=footnotesize]{subfig}

\usepackage{caption}

\usepackage{stfloats}
\usepackage{placeins}

\usepackage{enumitem}

\usepackage{multirow}

\newcommand{\thdr}[1]{\multicolumn{1}{c}{#1}}
\newcommand{\thdrl}[1]{\multicolumn{1}{c|}{#1}}

\usepackage{siunitx}
\DeclareSIUnit{\nothing}{\relax}

\usepackage{graphicx}
\usepackage{bm}
\usepackage{xcolor}

\SetCommentSty{algcommfont}

\usepackage{standalone}
\usepackage{tikz}
\usetikzlibrary{calc}
\usetikzlibrary{math}
\usetikzlibrary{decorations.pathreplacing}
\usetikzlibrary{arrows}
\usetikzlibrary{patterns}
\usetikzlibrary{positioning,arrows,arrows.meta}
\usetikzlibrary{shadows}

\definecolor{unblue}{rgb}{0.36, 0.57, 0.9}

\newcommand{\ttimes}{\raisebox{0.15ex}{{\scriptsize$\times$}}\xspace}

\theoremstyle{definition}

\newcommand{\sample}{\stackrel{\$}{\gets}}

\DeclareMathOperator\erf{erf}

\AtBeginDocument{%
  \providecommand\BibTeX{{%
    \normalfont B\kern-0.5em{\scshape i\kern-0.25em b}\kern-0.8em\TeX}}}

\copyrightyear{2023}
\acmYear{2023}
\setcopyright{acmlicensed}\acmConference[CCS '23]{Proceedings of the 2023 ACM SIGSAC Conference on Computer and Communications Security}{November 26--30, 2023}{Copenhagen, Denmark}
\acmBooktitle{Proceedings of the 2023 ACM SIGSAC Conference on Computer and Communications Security (CCS '23), November 26--30, 2023, Copenhagen, Denmark}
\acmPrice{15.00}
\acmDOI{10.1145/3576915.3623159}
\acmISBN{979-8-4007-0050-7/23/11}

\begin{document}

\title[FPT: a Fixed-Point Accelerator for Torus Fully Homomorphic Encryption]{FPT: a Fixed-Point Accelerator for\\Torus Fully Homomorphic Encryption}

\author{Michiel Van Beirendonck}
\orcid{0000-0002-5131-8030}
\email{michiel.vanbeirendonck@esat.kuleuven.be}
\affiliation{%
  \institution{COSIC, KU Leuven}
  \streetaddress{Kasteelpark Arenberg 10 - bus 2452}
  \city{Leuven}
  \country{Belgium}
  \postcode{3001}
}

\author{Jan-Pieter D'Anvers}
\orcid{0000-0001-9675-7988}
\email{janpieter.danvers@esat.kuleuven.be}
\affiliation{%
  \institution{COSIC, KU Leuven}
  \streetaddress{Kasteelpark Arenberg 10 - bus 2452}
  \city{Leuven}
  \country{Belgium}
  \postcode{3001}
}

\author{Furkan Turan}
\orcid{0000-0002-0045-7794}
\email{furkan.turan@esat.kuleuven.be}
\affiliation{%
  \institution{COSIC, KU Leuven}
  \streetaddress{Kasteelpark Arenberg 10 - bus 2452}
  \city{Leuven}
  \country{Belgium}
  \postcode{3001}
}

\author{Ingrid Verbauwhede}
\orcid{0000-0002-0879-076X}
\email{ingrid.verbauwhede@esat.kuleuven.be}
\affiliation{%
  \institution{COSIC, KU Leuven}
  \streetaddress{Kasteelpark Arenberg 10 - bus 2452}
  \city{Leuven}
  \country{Belgium}
  \postcode{3001}
}

\renewcommand{\shortauthors}{Van Beirendonck et al.}

\begin{abstract}

    Fully Homomorphic Encryption (FHE) is a technique that allows computation on encrypted data. It has the potential to drastically change privacy considerations in the cloud, but high computational and memory overheads are preventing its broad adoption. TFHE is a promising Torus-based FHE scheme that heavily relies on bootstrapping, the noise-removal tool invoked after each encrypted logical/arithmetical operation.

    We present FPT, a Fixed-Point FPGA accelerator for TFHE bootstrapping. FPT is the first hardware accelerator to heavily exploit the inherent noise present in FHE calculations. Instead of double or single-precision floating-point arithmetic, it implements TFHE bootstrapping entirely with approximate fixed-point arithmetic. Using an in-depth analysis of noise propagation in bootstrapping FFT computations, FPT is able to use noise-trimmed fixed-point representations that are up to 50\% smaller than prior implementations that prefer floating-point or integer FFTs.

    FPT is built as a streaming processor inspired by traditional streaming DSPs: it instantiates directly cascaded high-throughput computational stages, with minimal control logic and routing networks. We explore different throughput-balanced compositions of streaming kernels with a user-configurable streaming width in order to construct a full bootstrapping pipeline. Our proposed approach allows 100\% utilization of arithmetic units and requires only a small bootstrapping key cache, enabling an entirely compute-bound bootstrapping throughput of 1 BS / 35$\mu$s. This is in stark contrast to the established classical CPU approach to FHE bootstrapping acceleration, which is typically constrained by memory and bandwidth.

    FPT is fully implemented and evaluated as a bootstrapping FPGA kernel for an Alveo U280 datacenter accelerator card. FPT achieves two to three orders of magnitude higher bootstrapping throughput than existing CPU-based implementations, and 2.5\ttimes higher throughput compared to recent ASIC emulation experiments.

\end{abstract}



\settopmatter{printfolios=true}
\maketitle

\section{Introduction and Motivation}

Machine Learning (ML), driven by the availability of an abundance of data, has seen rapid advances in recent years \cite{DBLP:journals/csur/PouyanfarSYTTRS19}, leading to new applications from autonomous driving \cite{DBLP:journals/access/YurtseverLCT20} to medical diagnosis \cite{DBLP:journals/artmed/Kononenko01}. In many applications, ML models are developed by one party, who makes them available to users as a cloud service \cite{DBLP:journals/cacm/ArmbrustFGJKKLPRSZ10}. The deployment of such applications comes at the risk of privacy breaches, where the user data might be leaked, or IP theft, where users steal the ML model from the developing party \cite{DBLP:conf/eurosp/PapernotMSW18}.

\begin{figure}[tb!]
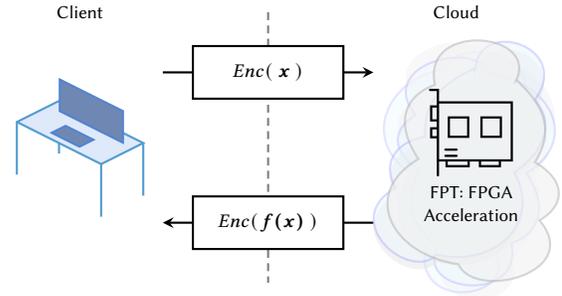

    \centering
    \includestandalone{figures/fhe}
    \caption{
        In FHE applications, critical computations occur on a cloud server over encrypted data. The server cannot decrypt the data, ensuring its privacy during computation. Decryption is solely possible by the client. Despite being slow, FHE enables previously impossible applications, at least impossible with the same level of security. FPT ---our proposed FPGA accelerator for cloud servers--- makes FHE practical by boosting computation speed.
    }
    \label{fig:FHE}
\end{figure}

The ``silver bullet'' solution to prevent the leakage of this data is to encrypt it with Fully Homomorphic Encryption (FHE)~\cite{rivest1978data,DBLP:conf/stoc/Gentry09}, which is a technique that allows one to compute on encrypted data. \cref{fig:FHE} illustrates a possible application of FHE to protect user data in an ML environment. In this scenario, a client wants to use an online-server-based ML service, without leaking any sensitive data. To this end, the client encrypts their data with FHE, before sending it to the cloud. The cloud service then computes an FHE program on the encrypted data without obtaining any information about the input and sends the (still encrypted) result back to the client. Only the client can finally decrypt and obtain the result.

The drawback of FHE is that it is at the moment still orders of magnitude slower than unencrypted calculations. The first algorithm to calculate an encrypted AND gate took up to 30 minutes to finish\cite{DBLP:conf/eurocrypt/GentryH11}. FHE schemes and algorithms have seen significant improvements in recent years, e.g. the recent TFHE scheme computes encrypted AND gates in only 13ms \cite{DBLP:conf/asiacrypt/ChillottiGGI16,DBLP:journals/joc/ChillottiGGI20} on a CPU. However, even with these improvements, it is not uncommon to still see slowdown factors of 10,000\ttimes compared to calculations on unencrypted data \cite{DBLP:journals/access/LeeKLCEDLLYKN22,DBLP:conf/aaai/HanHC019,DBLP:conf/cscml/ChillottiJP21}, which currently still prevents practical deployment of FHE in many applications.

To work around the speed limitations of FHE, designers have shifted their focus from general-purpose CPUs to more dedicated hardware implementations. Of these dedicated implementations, GPU-based FHE accelerators are easiest to develop, but they typically only provide modest speedups \cite{DBLP:conf/hpec/WangHCHS12,DBLP:journals/tches/JungKACL21,DBLP:conf/balkancryptsec/DaiS15,DBLP:journals/tches/BadawiVMA18}. At the other end of the spectrum, ASIC emulations in advanced technology nodes promise better FHE acceleration \cite{DBLP:conf/isca/KimKKJKRA22,DBLP:conf/micro/SamardzicFKDDP021,DBLP:conf/isca/SamardzicFKMGDE22,DBLP:journals/corr/abs-2205-14017,DBLP:conf/micro/KimLKSRKA22}. However, it can take years for these ASICs to be fabricated and become available \cite{DBLP:journals/corr/abs-2204-08742}, and they are typically specialized for a limited range of parameter sets. Finally, FPGA-based implementations can be developed more quickly than ASIC implementations, are flexible to change parameter sets, and can be readily deployed in FPGA-equipped cloud instances while boosting large speedups. As a result, they have been a popular target for FHE acceleration \cite{DBLP:conf/asplos/RiaziLPD20,DBLP:journals/tc/TuranRV20,DBLP:journals/corr/abs-2210-05476,DBLP:journals/corr/abs-2207-11872,DBLP:journals/tc/DorozOS15,DBLP:conf/ches/PoppelmannNPM15,DBLP:conf/hpca/RoyT0VV19}.

One costly operation in FHE calculations is bootstrapping. All currently available FHE schemes have an inherent noise that is increased with each operation. After a certain number of operations, this noise needs to be reduced to allow further calculations, which is done using this so-called bootstrap procedure.

Second-generation FHE schemes BFV \cite{DBLP:journals/iacr/FanV12}, BGV \cite{DBLP:journals/toct/BrakerskiGV14}, and their successor CKKS \cite{DBLP:conf/asiacrypt/CheonKKS17} -- sometimes called a fourth-generation FHE scheme -- have been the main focus of prior hardware accelerators. These schemes require bootstrapping only after a certain number of operations. For these schemes, bootstrapping is a complex algorithm that requires large data caches \cite{DBLP:journals/corr/abs-2112-06396} and exhibits low arithmetic intensity, and essentially all prior architectures that support bootstrapping have hit the off-chip memory-bandwidth wall \cite{DBLP:conf/isca/KimKKJKRA22,DBLP:conf/micro/SamardzicFKDDP021}.

Third-generation schemes like FHEW \cite{DBLP:conf/eurocrypt/DucasM15} and its successor Torus FHE (TFHE)~\cite{DBLP:conf/asiacrypt/ChillottiGGI16,DBLP:journals/joc/ChillottiGGI20} have revisited the bootstrapping approach, making it cheaper but inherently linked to homomorphic calculations. In these schemes, most of the homomorphic operations require an immediate bootstrap of the ciphertext. Moreover, bootstrapping in TFHE is a versatile tool, which can additionally be ``programmed'' with an arbitrary function that is applied to the ciphertext, e.g. non-linear activation functions in ML neural networks \cite{DBLP:conf/cscml/ChillottiJP21}. This approach is called Programmable Bootstrapping (PBS) and it constitutes the main cost of TFHE homomorphic calculations. Taking up to 99.9\% of an encrypted circuit computation (\cref{sec:gol}), PBS is a prime target for high-throughput hardware acceleration of TFHE.

In this work, we propose FPT, an FPGA-based accelerator for TFHE Programmable Bootstrapping. FPT achieves a significant speedup over the previous state-of-the-art, which is attributable to two major contributions:

\begin{enumerate}[leftmargin=*]
    \item FPT's microarchitecture is built as a streaming processor, challenging the established classical CPU approach to FHE bootstrapping accelerators. Inspired by traditional streaming DSPs, FPT instantiates high-throughput computational stages that are directly cascaded, with simplified control logic and routing networks. FPT's streaming approach allows 100\% utilization of arithmetic units during bootstrapping, including tool-generated high-radix and heavily optimized negacyclic FFT units with user-configurable streaming widths. Our streaming architecture is discussed in \cref{sec:architecture}.

    \item FPT (\textbf{F}ixed-\textbf{P}oint \textbf{T}FHE) is the first hardware accelerator to extensively optimize the representation of intermediate variables. TFHE PBS is dominated by FFT calculations, which work on irrational (complex) numbers and need to be implemented with sufficient accuracy. Instead of using double floating-point arithmetic or large integers as in previous works, FPT implements PBS entirely with compact fixed-point arithmetic. We analyze in-depth the noise due to the compact fixed-point representation that we use inside PBS, and we match it to the noise that is natively present in FHE. Through this analysis, FPT is able to use fixed-point representations that are up to 50\% smaller than prior implementations using floating-point or integer FFTs. In turn, these 50\% smaller fixed-point representations enable up to 80\% smaller FFT kernels. Our fixed-point analysis is discussed in \cref{sec:fixedpointparams}.
\end{enumerate}

\noindent FPT shows, for the first time, that PBS can remain entirely compute-bound with only small bootstrapping key data caches. FPT achieves a massive PBS throughput of 1 PBS / 35$\mu$s, which requires only modest off-chip memory bandwidth, and is entirely bound by the logic resources on our target AMD Alveo U280 FPGA. This represents almost three orders of magnitude speedup over the popular TFHE software library CONCRETE \cite{chillotti2020concrete} on an Intel Xeon Silver 4208 CPU at \SI{2.1}{\giga\hertz}, a factor 7.1\ttimes speedup over a concurrently-developed FPGA architecture \cite{DBLP:conf/hpec/YeKP22}, and a factor 2.5\ttimes speedup over recent \SI{16}{\nano\meter} ASIC emulation experiments \cite{DBLP:conf/dac/0001LJ22}.

\section{Background}

This section gives an intuitive idea of the workings of TFHE, with a focus on the Programmable Bootstrapping step that is accelerated by FPT. We refer the reader to~\cite{DBLP:conf/asiacrypt/ChillottiGGI16,DBLP:journals/joc/ChillottiGGI20,DBLP:journals/tches/Joye22} for a more in-depth overview of TFHE.

\subsection{Torus Fully Homomorphic Encryption}
Torus Fully Homomorphic Encryption (TFHE) is a homomorphic encryption scheme based on the Learning With Errors (LWE) problem. It operates on elements that are defined over the real Torus $\mathbb{T} = \mathbb{R}/\mathbb{Z}$, i.e. the set $[0, 1)$ of real numbers modulo $1$. In practice, Torus elements are discretized as 32-bit or 64-bit integers.

A TFHE ciphertext can be constructed by combining three elements: a secret vector $s$ with $n$ coefficients following a uniform binary distribution $s \sample \mathcal{U}(\mathbb{B}^n)$, a public vector $a \sample \mathcal{U}(\mathbb{T}^n)$ sampled from a uniform distribution, and a small error $e \sample \chi$ from a small distribution $\chi(\mathbb{T})$. A message $\mu \in \mathbb{T}$ can be encrypted as a tuple: $c = (a, b = a \cdot s + e + \mu) \in \mathbb{T}^{n+1}$. Using the secret $s$, one can decrypt the ciphertext back into (a noisy version of) the message by computing $b - a \cdot s = \mu + e $. This type of ciphertext is called a Torus LWE (TLWE) ciphertext.

TFHE additionally describes two variant ciphertexts: First, a generalized version (TGLWE), where $e$ and $\mu$ are polynomials in $\mathbb{T}_{N}[X] = \mathbb{T}[X]/(X^{N}+1)$, and where $a$ and $s$ are vectors of polynomials of the form $\mathbb{T}_{N}[X]^{k}$. The TGLWE ciphertext is then similarly formed as a tuple: $c = (a, b = a \cdot s + e + \mu) \in \mathbb{T}_N[X]^{k+1}$. The second variant is a generalized version of a GSW \cite{DBLP:conf/crypto/GentrySW13} ciphertext (TGGSW), which is essentially a matrix where each row is a TGLWE ciphertext: $c \in \mathbb{T}_{N}[X]^{(k+1)l \times (k+1)}$.

The reason for defining TGLWE and TGGSW ciphertexts is that they permit a homomorphic multiplication:
\begin{equation*}
    \text{TGLWE}(\mu_1) \boxdot \text{TGGSW}(\mu_2) = \text{TGLWE}(\mu_1 \cdot \mu_2),
\end{equation*} known as the \emph{External Product} ($\boxdot$). First, it decomposes each of the polynomials in the TGLWE ciphertext into $l$ polynomials of $\beta$ bits, an operation termed gadget decomposition. Next, the decomposed TGLWE ciphertext and TGGSW are multiplied in a $(k+1)l-$vector times $(k+1)l \times (k+1)$-matrix product where the elements of this vector and matrix are polynomials in $\mathbb{T}_{N}[X]$. The output is again a TGLWE ciphertext encrypting $\mu_1 \cdot \mu_2$.

\subsection{Programmable Bootstrapping}
The main goal of bootstrapping is to reduce the noise in the ciphertext. One way to reduce the ciphertext noise would be to decrypt the ciphertext, after which the noise can be suppressed, but this would not be secure. Bootstrapping does in essence decrypt the ciphertext, but for security reasons this operation is performed, homomorphically, inside the encrypted domain. This means that one wants to homomorphically compute $b - a \cdot s = e + \mu$, and more specifically, as it is ``programmable'' bootstrapping, one wants to additionally compute a function $f(\mu)$ on the data.

To achieve this programmable bootstrapping, one first sets a ``test'' polynomial $F = \sum_{i=0}^{N-1} f(i) \cdot X^i \in \mathbb{T}_{N}[X]$ that encodes $N$ relevant values of the function $f$. This polynomial is then rotated with $b - a \cdot s$ positions by calculating $F \cdot X^{- (b - a \cdot s)}$, after which the output to the function can be found on the first position of the resulting polynomial. However, all of these calculations should be done without revealing the value of $s$.

The high-level idea of how to achieve this is to first rewrite the above expression as follows:
\begin{align}
     & F \cdot X^{- (b - a \cdot s)} = F \cdot X^{- b} \cdot \prod_{i=1}^{n} X^{a_i \cdot s_i}.
\end{align}
This expression can be calculated iteratively. Starting with the polynomial $ACC = F \cdot X^{- b}$, one iteratively calculates:
\begin{align}
    ACC \gets ACC \cdot X^{a_i \cdot s_i},
\end{align}
which can be further rewritten, using the fact that $s_i$ is either zero or one, to:
\begin{align}
    ACC \gets (ACC \cdot X^{a_i} - ACC ) \cdot s_i + ACC.
\end{align}

\noindent However, as we cannot reveal $s_i$, we encode the $s_i$ value in a TGGSW ciphertext $BK_i$, and the $ACC$ value in a TGLWE ciphertext, after which the expression becomes:
\begin{align}
    \label{eq:cmux}
    ACC \gets (ACC \cdot X^{a_i} - ACC ) \boxdot BK_i + ACC,
\end{align}
using the homomorphic multiplication operation $\boxdot$. \cref{eq:cmux} homomorphically multiplexes on the secret value $s_i$, and is known as the Controlled MUX (CMUX).

Collectively, the different TGGSW ciphertexts $BK_1, \dots, BK_{n}$, each encrypting one secret coefficient $s_1, \cdots, s_{n}$, are known as the bootstrapping key. The result of the operations described above is a TGLWE accumulator $ACC$ which is ``blindly'' rotated with a secret amount of $b-a \cdot s$ positions, from which the output TLWE ciphertext can be straightforwardly extracted. The computations during PBS are given in \cref{alg:PBS}.

\begin{algorithm}
    \footnotesize
    \SetNoFillComment
    \SetKwInOut{Input}{input}\SetKwInOut{Output}{output}
    \SetKwComment{Comment}{/* }{ */}
    
    \DontPrintSemicolon
    
    \caption{TFHE's Programmable Bootstrapping}\label{alg:PBS}
    \tcp*[l]{TLWE Ciphertext}
    \Input{${c}_{in} = (a_1,\ldots,a_n, b) \in \mathbb{T}^{n+1}$}
    \tcp*[l]{TGGSW Bootstrapping Key}
    \Input{${BK} = (BK_1,\ldots,BK_n) \in \mathbb{T}_{N}[X]^{n \times (k + 1)l \times (k+1)}$}
    \tcp*[l]{TGLWE Test Polynomial LUT}
    \Input{$F \in \mathbb{T}_{N}[X]^{(k+1)}$} 
    \tcp*[l]{TLWE Ciphertext}
    \Output{${c}_{out} \in \mathbb{T}^{kN+1}$}
    
    \BlankLine
    \tcp*[l]{Test Polynomial LUT}
    $ACC \gets F \cdot X^{-b}$  \label{line:pbsinit} %
    \BlankLine
    
    \tcp*[l]{Blind Rotation}
    \For{$i \gets 1$ \KwTo $n$ \label{line:pbsblindrotate}}{
        \tcp*[l]{CMUX}
        $ACC \gets  (ACC \cdot X^{\lfloor \frac{2Na_i}{q} \rceil}  - ACC) \boxdot BK_i + ACC$ \label{line:pbscmux} %
    }
    \KwRet{$c_{out} = \texttt{SampleExtract}(ACC)$ \label{line:pbssamplex}}
    \end{algorithm}

FPT implements two parameter sets of TFHE, given in \cref{tab:params}. Parameter Set I is a parameter set used by the CONCRETE Boolean library with 128-bit security \cite{chillotti2020concrete}. Parameter Set II is a 110-bit security parameter set that has previously been employed for benchmarking purposes, allowing a direct comparison of FPT with prior work \cite{DBLP:journals/joc/ChillottiGGI20,zama_2022}.

\begin{table}
    \centering
    \captionsetup{singlelinecheck=off}
    \caption[Parameter Sets]{Parameter Sets:\\
    I is used by the CONCRETE Boolean library \cite{chillotti2020concrete}.\\
    II is popular for benchmarking purposes \cite{DBLP:journals/joc/ChillottiGGI20,zama_2022}.}
    \begin{tabular}{llS[table-format=4.0]S[table-format=4.0]}
  \toprule 
  \multicolumn{2}{l}{Parameter Set} & I         & II        \\
  \midrule
  Security Level   &                & {128-bit} & {110-bit} \\
  \arrayrulecolor{gray!50}\midrule
  TLWE dimension   & n              & 586       & 500       \\
  TGLWE dimension  & k              & 2         & 1         \\
  Polynomial size  & N              & 512       & 1024      \\
  Decomp. Base Log & $\beta$        & 8         & 10        \\
  Decomp. Level    & $l$            & 2         & 2         \\
  \arrayrulecolor{black}\bottomrule
\end{tabular}
    \label{tab:params}
\end{table}

\subsection{FFT polynomial multiplications}
\label{sec:negacyclicfft}

As seen in \cref{alg:PBS}, the TFHE programmable bootstrapping mainly consists of iterative calculation of the external product $\boxdot$, which is a vector-matrix multiplication where the elements are large polynomials of order $N$. Underlying polynomial multiplications, therefore, dominate the bootstrapping.

A schoolbook approach to polynomial multiplication would result in a computational complexity $O(N^2)$. However, utilizing the convolution theorem, the FFT can be used to compute these polynomial multiplications in time $O(N \log(N))$, as the multiplication of polynomials modulo $X^N-1$ corresponds to a cyclic convolution of the input vectors. FHE schemes, however, need polynomial multiplications modulo $X^N+1$, requiring \emph{negacyclic} FFTs to compute negative-wrapped convolutions. This negacyclic convolution has a period $2N$, and thus a straightforward implementation would require $2N$ size FFTs. The cost of the negacyclic FFT on real input data can be reduced using two techniques.

The fact that the FFT computes on complex numbers offers the first opportunity for optimization. Since the input polynomials are purely real and have an imaginary component equal to zero, real-to-complex (r2c) optimized FFTs can be used, which achieve roughly a factor of two improvements in speed and memory usage \cite{1386650}. This is the approach taken by the TFHE and FHEW software libraries, which compute size-2N r2c FFTs.

A second possible optimization is that negacyclic FFTs, which would have a period and size of $2N$, can be computed instead as a regular FFT with period and size $N$ by using a ``twisting'' preprocessing step~\cite{DBLP:books/aw/AhoHU74}. During twisting, the coefficients of the input polynomial $a$ are multiplied with the powers of the $2N$-th root of unity $\psi = \omega_{2N}$,
\begin{equation}
    \hat{a} = (a[0], \psi a[1], \ldots, \psi^{N-1}a[N-1]).
\end{equation}
After twisting, one can perform multiplication using a regular cyclic FFT on $\hat{a}$, halving the required FFT size to $N$.

While both optimizations are well-known individually, it is less straightforward to combine them. Intuitively, the twisting step is incompatible with the r2c optimization, because it will make the polynomial complex.

We use a third, but not-so-well-known technique from NuFHE \cite{nufhe} based on the tangent FFT \cite{DBLP:conf/aaecc/Bernstein07}. The crux of this method is to ``fold'' polynomial coefficients $a[i]$ and $a[i+N/2]$ into a complex number $a[i]+ja[i+N/2]$ before applying the twisting step and subsequent cyclic size-$N/2$ FFT. This quarters the size of the FFT required from the original naive size-$2N$ FFT. We adopt this technique in FPT and use FFTs of size $N/2=256$ and $N/2=512$ for Parameters Sets I and II (\cref{tab:params}), respectively.

\section{FPT Microarchitecture}
\label{sec:architecture}

In this section, we discuss FPT's microarchitecture. First, we describe how FPT's architecture is designed as a streaming processor targeting maximum throughput. Next, we detail a batch bootstrapping technique, which significantly reduces FPT's on-chip caches and off-chip bandwidth. Finally, we present balanced implementations of the various computational stages, which enable 100\% utilization of the arithmetic units during FPT's bootstrapping operation.

\subsection{Streaming Processor}

FHE accelerators for second-generation schemes have mostly been built after a classical CPU architecture \cite{DBLP:conf/micro/SamardzicFKDDP021,DBLP:journals/corr/abs-2205-14017,DBLP:conf/micro/KimLKSRKA22}. They include a control unit that executes an instruction set, together with a set of arithmetic Processing Elements (PEs) that support different operations, e.g. ciphertext multiplication, key-switching, or bootstrapping.
Different operations utilize different PEs, requiring careful profiling of FHE programs to balance PE relative throughputs and utilization \cite{DBLP:conf/isca/KimKKJKRA22, DBLP:conf/isca/SamardzicFKMGDE22}.

These accelerators are often memory-bound during bootstrapping, and in order to keep a high utilization level of PEs, an increasing focus is spent on optimizing the memory hierarchy, often including a multi-layer on-chip memory hierarchy with a large ciphertext register file at the lowest level.

FPT challenges this established classical CPU approach to FHE bootstrapping acceleration, and instead adopts a microarchitecture that is inspired by streaming Digital Signal Processors (DSPs). Data flows naturally through FPT's wide and directly cascaded computational stages, with simplified hard-wired routing paths and without complicated control logic. During FPT's bootstrapping operation, utilization of arithmetic units is 100\%.

\begin{figure*}
    \centering
    \includegraphics[width=0.95\linewidth]{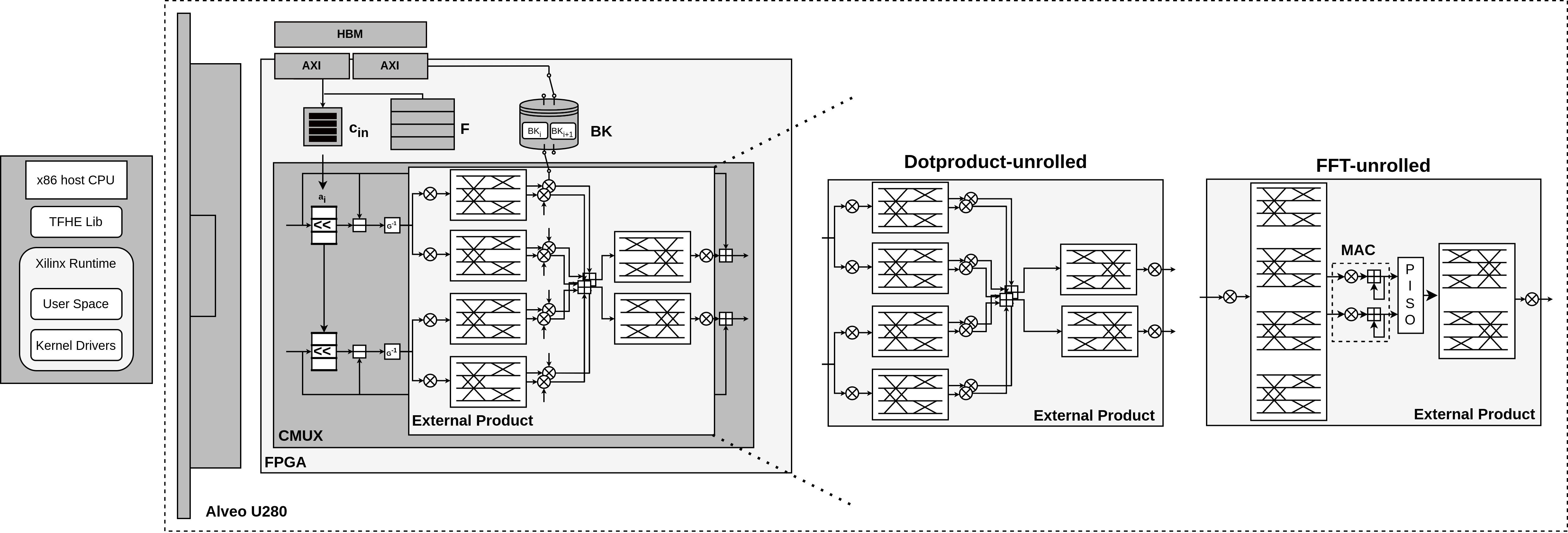}
    \caption{FPT's microarchitecture. FPT instantiates only a single PE, the CMUX PE. The CMUX is built with wide, directly cascaded datapaths, targeting massive throughput. In light grey are illustrated two throughput-balanced architectures for the external product (with $k=1$, $l=2$): dotproduct-unrolled (left) and FFT-unrolled (right). Host-FPGA communication includes three different interfaces: an input ciphertext FIFO, a ping-pong bootstrapping-key buffer, and a test polynomial $F$ SRAM.}
    \label{fig:FPT}
\end{figure*}

As illustrated in \cref{fig:FPT}, FPT defines only a single fixed PE, the CMUX PE, and instantiates only a single instance of this PE with wide datapaths and massive throughput. Taking advantage of the regular structure of TFHE's PBS, consisting of $n$ repeated CMUX iterations, this single high-throughput PE suffices to run PBS to completion. The CMUX PE computes a single PBS CMUX iteration, after which its datapath output hard-wires back into its datapath input.

Internally, the CMUX PE computes a fixed sequence of monomial multiplication, gadget decomposition, and polynomial multiply-add operations of the external product. Rather than dividing the CMUX into sub-PEs that are sequenced to run from a register file, FPT builds the CMUX with directly cascaded computational stages. Stages are throughput-balanced in the most conceivably simple way: each stage operates at the same throughput and processes a number of polynomial coefficients per clock cycle that we call the \emph{streaming width}. Stages are interconnected in a simple fixed pipeline with static latency, avoiding complicated control logic and simplifying routing paths.

FPT is built to achieve maximum PBS throughput. As a general trend that we will detail later (\cref{fig:sgen-synth2}), the Throughput/Area (TP/A) of computational stages increases together with the streaming width. This motivates FPT to instantiate only a single wide CMUX PE with high streaming width, as opposed to many CMUX PEs with smaller streaming widths.

In summary, FPT's CMUX architecture enables massive PBS throughput by more closely resembling the architecture of a streaming Digital Signal Processor (DSP), rather than the classical CPU architecture employed by prior FHE processors.

\subsection{Batch Bootstrapping}
\label{subsec:batchbootstrapping}

TFHE bootstrapping requires two major inputs: the input ciphertext coefficients $a_1, \dots, a_n$ and the bootstrapping keys $BK_1, \dots, BK_n$. Each iteration of the CMUX PE requires one element of both. The ciphertext coefficients $a_i$ are relatively small in size; therefore, they are easy to accommodate. In contrast, a bootstrapping key coefficient $BK_i \in \mathbb{T}_{N}[X]^{(k + 1)l \times (k+1)}$ is a large matrix of up to tens of kBs. Since the full $BK$ is typically too large to fit entirely on-chip, the $BK_i$ must be loaded from off-chip memory for every iteration. However, at high CMUX throughput levels, the required bandwidth for $BK_i$ could easily exceed 1.0~TB/s. This is larger even than the bandwidth of HBM, and thus poses a memory bottleneck.

We propose a method, termed \emph{batch bootstrapping}, to amortize loading the bootstrapping key for each iteration. The result is that FPT can operate entirely compute-bound, with modest off-chip bandwidth and small on-chip caches. In contrast, prior FHE processors that supported bootstrapping of second-generation schemes were often bottlenecked by the required memory bandwidth \cite{DBLP:conf/isca/KimKKJKRA22,DBLP:conf/micro/SamardzicFKDDP021}. In fact, a recent architectural analysis of bootstrapping \cite{DBLP:journals/corr/abs-2112-06396} found that it exhibits low arithmetic intensity and requires large caches. Their conclusion was that FHE processors only benefit marginally from bespoke high-throughput arithmetic units. With our design, we show that the situation can be very different for TFHE's PBS.

In FPT, we solve the memory bottleneck problem as follows: First, due to internal pipelining, the latency of the CMUX will be much larger than its throughput. To operate at peak throughput, FPT processes multiple ciphertexts to keep its CMUX pipeline stages full. Next, we enforce that the different ciphertexts processed concurrently in the CMUX's pipeline stages arrive in a single batch of size $b$, encrypted under the same $BK$. This ensures that these ciphertexts are at the same CMUX iteration, and as a result, all require the exact same input coefficient $BK_i$.

Batch bootstrapping then proceeds as follows: We instantiate a simple BRAM ping-pong buffer that holds two coefficients of $BK$. The CMUX reads $BK_{i}$ from one half with the required bandwidth of 1.0~TB/s, while the off-chip memory fills $BK_{i+1}$ inside the other half with a bandwidth of $1.0/b$~TB/s. In a technique similar to C-slow retiming \cite{DBLP:journals/algorithmica/LeisersonS91}, we can arbitrarily increase the batch size $b$ by introducing more pipeline registers within the CMUX, without throughput penalty. With a batch size of $b = 16$, already the required bandwidth can be supplied by DDR4 instead of HBM. Such a small batch size is typically easy to accommodate for FHE applications, e.g. homomorphic neural networks feature hundreds to thousands of parallel ciphertext inputs \cite{DBLP:conf/cscml/ChillottiJP21}.

Our simple but crucial batch bootstrapping technique exploits locality of reference to decouple the on-chip bandwidth from the off-chip bandwidth. As a result, in our architecture, TFHE's PBS is entirely compute-bound with only kB-size caches, not larger than the size of two coefficients of the bootstrapping key.

\subsection{Balancing the External Product}

The external product $(\boxdot)$, computing a vector-matrix negacyclic polynomial product, represents the bulk of the CMUX logic. As discussed before, the polynomial multiplications are performed using an FFT, and thus the $(\boxdot)$ operations include forward and inverse negacylic FFT computations, and pointwise dot-products with $BK_i$ (the bootstrapping key $BK_i$ is already in the FFT domain).

In a streaming architecture, it is important to balance throughputs of processing elements, which is not trivial as the external product includes $(k+1)l$ forward FFTs, but only $(k+1)$ inverse FFT operations. We explore two different throughput-balanced architectures for the external product as shown in light-grey in \cref{fig:FPT}: a dotproduct-unrolled architecture (left) and an FFT-unrolled architecture (right).

The dotproduct-unrolled architecture (left) represents the more obvious choice for parallelism, where we instantiate $l$ times more FFT kernels compared to IFFT kernels. With the FFT-unrolled architecture on the right, we make a more unconventional choice: we balance throughputs by instantiating the FFT with $l$ times the streaming width of the IFFT. These two architectural trade-offs can be understood as exploiting different types of ``loop unrolling'' inside the external product. On the left, we first loop-unroll the dot-product before unrolling the FFT, while on the right, we loop-unroll the FFT maximally.

The drawback of the FFT-unrolled architecture is that it is more complex than the dotproduct-unrolled one. First, multiply-add operations must be replaced by MACs, since polynomial coefficients that must be added are now spaced temporally over different clock cycles. Second, the inverse FFT can only start processing once a full MAC has been completed, requiring a Parallel-In Serial-Out (PISO) block that double-buffers the MAC output and matches throughputs. Third and most importantly, FFT blocks can be challenging to unroll and implement for arbitrary throughputs, and supporting two FFT blocks with differing throughputs requires non-negligible extra engineering effort.

The main advantage of the relatively unconventional FFT-unrolled architecture is that it features fewer FFT kernels, which feature higher streaming widths. As we will detail in the next section, this favors the general (and often-neglected) trend of pipelined FFTs, which typically feature significantly higher TP/A as the streaming width increases. At the most extreme end, a fully parallel FFT is a circuit with only constant multiplications and fixed routing paths, featuring up to 300\% more throughput per DSP or per LUT on our target FPGA (\cref{fig:sgen-synth2}). FPT alleviates the extra engineering effort and extra complexity of the FFT-unrolled architecture, by extending and optimizing an existing FFT generator tool to support negacylic FFTs.

\subsection{Streaming Negacylic FFTs}
\label{sec:streamingFFTs}

State-of-the-art FHE processors have implemented mostly \emph{iterative} FFTs or NTTs that process polynomials in multiple passes \cite{DBLP:journals/corr/abs-2210-05476,DBLP:journals/corr/abs-2207-11872,DBLP:conf/asplos/RiaziLPD20,DBLP:journals/corr/abs-2205-14017}. In these architectures, it can be difficult to support arbitrary throughputs, as memory conflicts arise when each pass requires data at different strides. Instead, FPT instantiates \emph{continuous-flow pipelined} FFTs that naturally support a streaming architecture. Pipelined FFT architectures consist of $log(N)$ stages that are connected in series. The main advantage of these architectures is that they process a continuous flow of data, which lends itself well to a fully streaming external product design.

There are many pipelined FFT architectures that target high-throughput and support arbitrary streaming widths, and we refer to \cite{Garrido2021} for a recent survey. Generally, pipelined FFTs cascade two types of units: first, the well-known butterflies with complex twiddle factor multipliers, and, second, shuffling circuits that compute stride permutations. Pipelined FFTs feature a large design space, with different possible overall architectures, area/precision trade-offs in computing twiddle factor ``rotations'', varying radix structures that determine which twiddle factors appear at which stages, and more. As such, they are an excellent target for tool-generated circuits, and we follow this approach for FPT.

Several FFT generator tools have been proposed in the literature. Some IP cores do not offer the massive parallelism and arbitrary streaming widths that we target for FPT \cite{fftlogicore,dblclkfft}. At the other end of the spectrum, a recent generator \cite{DBLP:journals/tcas/GarridoMK19} built on top of FloPoCo \cite{DBLP:journals/dt/DinechinP11} can only generate fully-parallel FFTs, instead of supporting arbitrary streaming widths. We synthesized at different streaming widths the High-Level Synthesis (HLS) Super Sample Rate (SSR) FFTs included in the Vitis DSP libraries of AMD \cite{vitislib}, but found that they are outperformed by the RTL Verilog FFTs generated by the Spiral FFT IP Core generator \cite{DBLP:journals/todaes/MilderFHP12}. Unfortunately, Spiral is not open-source and offers only a web interface towards its generated RTL \cite{spiralweb}.

Eventually, we settled on SGen\cite{DBLP:conf/fpl/SerreP18,DBLP:journals/trets/SerreP20,DBLP:conf/arith/SerreP19} as the FFT generator tool that provided the necessary configurability, extensibility, and performance we targeted for FPT. SGen is an open-source generator implemented in Scala and employs concepts introduced in Spiral. It generates arbitrary-streaming-width FFTs through four Intermediate Representations (IRs) with different levels of optimization: an algorithm-level representation SPL, a streaming-block-level representation Streaming-SPL, an acyclic streaming IR, and an RTL-level IR. Apart from the streaming width, SGen features a configurable FFT point size, radix, and hardware arithmetic representations such as fixed-point, IEEE754 floating-point, or FloPoCo floating-point.

Most importantly, SGen is fully open-source and extensible, which we make heavy use of to generate streaming FFTs for FPT. First, we have extended SGen with operators for the forward and inverse twisting step, necessary to support negacyclic FFTs (\cref{sec:negacyclicfft}). Next, we have implemented a set of optimizations aimed at higher precision and better TP/A. In this category, first, we have extended SGen with radix-$2^k$ structures \cite{DBLP:conf/ipps/HeT96,DBLP:journals/tvlsi/GarridoGSG13}, finding that radix-$2^4$ FFTs are on average 10\% smaller than SGen-generated radix-4 or radix-16 FFTs. Second, we replace schoolbook complex multiplication in SGen, requiring 4 real multiplies and 2 real additions, with a variant of Karatsuba multiplications that is sometimes attributed to Gauss: \begin{equation}
    \begin{split}
        X+jY &= (A+jB) \cdot (C+jD) \\
        Z &=  C \cdot (A-B) \\
        X &=  (C-D) \cdot B + Z\\
        Y &=  (C+D) \cdot A - Z
    \end{split}
    \label{eq:gauss}
\end{equation} By pre-computing $C-D$ and $C+D$ for the constant twiddle factors, this multiplication requires only 3 real multiplies and 3 adds, saving scarce FPGA DSP units.

Third, we decouple the twiddle bit-width from the input bit-width. On one hand, this allows us to take advantage of the asymmetric 27$\times$18 multipliers found in FPGA DSP blocks, while at the same time, it has been found that twiddles can be quantized with approximately four fewer bits without affecting output noise \cite{DBLP:journals/vlsi/CortesVZIS08,DBLP:journals/tsp/ChangN08}. Finally, as data grows throughout the FFT stages, it must initially be padded with zeros to prevent overflows. We have extended SGen with a scaling schedule, that instead divides the data by two whenever the most-significant bit must grow. Since the least-significant bits have mostly accumulated noise \cite{1162035}, scaling increases the precision for fixed input bit-width. Adding a scaling schedule allows us, on average, to use FFTs with 2-bit smaller fixed-point intermediate variables while meeting the same precision targets, which proves crucial to efficiently map multipliers to DSP units as will be detailed later in \cref{sec:fixedpointparams}, \cref{fig:areavswidth}.

\begin{figure}
    \centering
    \subfloat[\label{fig:sgen-synth}]{\includegraphics[width=\linewidth]{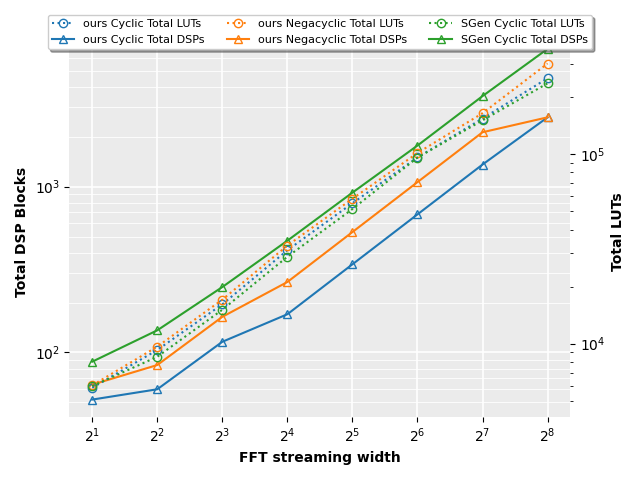}}  \\
    \subfloat[\label{fig:sgen-synth2}]{\includegraphics[width=\linewidth]{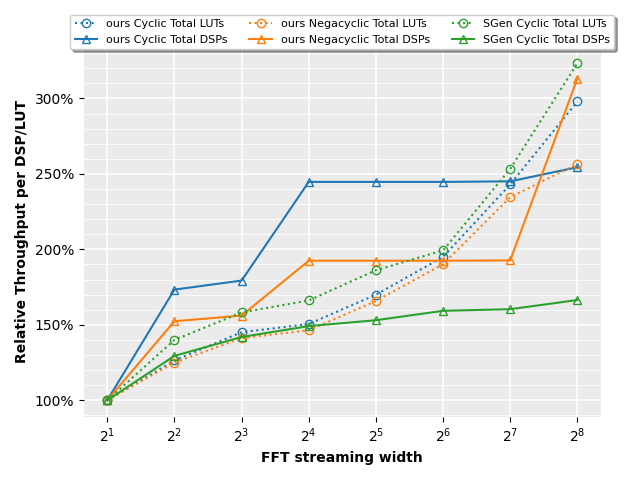}}
    \caption{FPGA resource utilization (a) and throughput / resource utilization (b) of a size-256 FFT at different streaming widths. At iso-precision, SGen FFTs use 31-bit intermediate variables without scaling schedule, and our FFTs use 29-bit intermediate variables with scaling.
    }
\end{figure}

Figure \ref{fig:sgen-synth} illustrates the resource usage of negacyclic size-256 FFTs produced by our optimized variant of SGen at different streaming widths. To quantize our improvements over SGen, we also add cyclic FFTs both with and without our introduced changes to the tool. Our changes result in significantly fewer logic resources: over 60\% fewer DSP blocks are utilized while keeping LUTs comparable. As DSP blocks are the main limiting resource for FPT (\cref{tab:utilization}), our optimizations are a key enabler to building FPT with high streaming widths.

Figure \ref{fig:sgen-synth2} illustrates our main motivation to propose the FFT-unrolled architecture for the external product. We plot the relative throughput per area unit (DSPs or LUTs) of tool-generated FFTs for different streaming widths. The trend is clear: FFTs with higher streaming widths feature up to 300\% more throughput per DSP or per LUT. Intuitively, as the streaming width increases, FFTs can take more advantage of the native strengths of hardware circuits. First, shuffling circuits with MUXes and storage blocks are replaced with fixed routing paths. Second, twiddle factor multipliers can be specialized to the specific set of twiddles they need to handle, taking advantage of optimized algorithms for Single- or Multiple Constant Multiplication (SCM, MCM).

\subsection{Other operations}

Compared to the external product and its streaming FFTs, the remainder of the CMUX (\cref{fig:FPT}, dark grey) represents mostly simple circuitry: additions, subtractions, gadget decomposition, and monomial multiplication. Whereas the first three can be streamed straightforwardly, monomial multiplication requires special treatment.

Monomial multiplication multiplies the accumulator $ACC$ with the ciphertext-dependent monomial $X^{\lfloor \frac{2Na_i}{q} \rceil}$. Its effect is to rotate the polynomials of $ACC$ by $\lfloor \frac{2Na_i}{q} \rceil$, and additionally negate those coefficients that wrap around. First, we truncate $\frac{2Na_i}{q}$ already in software to limit host-FPGA bandwidth. Next, an efficient architecture for monomial multiplication is a coefficient-wise barrel shifter in $log(N)$ stages.

We define two approaches to stream this operation; coefficient-wise and bitwise streaming. In coefficient-wise streaming, different polynomial coefficients are spaced temporally over different clock cycles. In bitwise streaming, all coefficients arrive in parallel within the same clock cycle; instead, we divide different bit chunks of each coefficient over different clock cycles. One can then make a simple observation: a rotation is a difficult permutation to stream coefficient-wise, as it must interchange coefficients that are spaced over different clock cycles. However, bitwise streaming is trivial, as it boils down to simply rotating all the individual bit-chunks. Therefore, we add stream-reordering blocks that switch a polynomial from coefficient-wise streaming to bitwise streaming and vice versa. At the same time, we merge the stream-reordering with the folding operation of the negacyclic FFT, which packs coefficients $a[i]$ and $a[i+N/2]$. The reordering block can be implemented at full throughput either in an R/W memory block or with a simple series of registers and MUXes.

Signed gadget decomposition involves taking unsigned 32-bit coefficients, and decomposing them into $l$ signed coefficients of $\beta$ bits. In hardware, this involves a simple reinterpretation of the bits and conditional subtraction. We merge this logic at the output of monomial multiplication to take advantage of LUT packing. In the bit-wise streamed representation, these operations must track the propagating carries in flip-flops.

Gadget decomposition is approximate, e.g., for Parameter Set I, $l \cdot \beta = 16$-bit $< 32$-bit. Contrarily to software implementations, FPT employs a CMUX datapath that is natively adjusted to approximate gadget decomposition. We discard bits prematurely that would later be rounded, allowing us to stick to a native 16-bit datapath, rather than growing back to 32-bit outside of the external product.

\section{Compact fixed-point representation}
\label{sec:fixedpointparams}
FFT calculations involve irrational (complex) numbers, and approximation errors arise when those numbers are represented with finite precision during computation.
However, if enough precision is used, implementations of TFHE tolerate these approximation errors. More specifically, one typically aims for the total approximation error to be lower than the noise inherently present in the FHE calculations. By selecting such parameters one can ensure that the error probability of the FHE calculations under full precision is arbitrarily close to the error probability of our calculations under finite precision, while obtaining a significant boost in performance.

It is important to note that using finite precision computations has no impact on the security of the cryptographic scheme. This can be easily seen by the fact that the server does not know the secret key and thus only performs public operations.

Floating-point number calculations are well supported on a CPU with single or double precision. Hence, they become the typical representation of choice for software designers. In the case of TFHE, the CPU and GPU implementations are restricted to double-precision floating-point FFTs, because single-precision FFTs were found to introduce too much noise to guarantee successful decryption of bootstrapped ciphertexts \cite{DBLP:journals/joc/ChillottiGGI20}.

In dedicated hardware implementations, FPUs are not natively available and are costly to include. To simplify the implementation and the analysis of the approximation error, some prior implementations opted to change the scheme to work with a prime modulus instead of a power-of-two modulus \cite{DBLP:conf/hpec/YeKP22,nufhe}, allowing the use of exact NTTs instead of approximate FFTs for polynomial multiplication. The downside of this approach is that one needs to include costly modular reduction units.

FPT is the first TFHE accelerator to instead utilize fixed-point calculations, which avoids the costly implementation of FPUs or modular reduction units. Moreover, instead of initializing very large fixed-point calculations to guarantee sufficient accuracy, we conduct an in-depth analysis that optimizes the fixed-point bit width to be just large enough so that the approximation noise is smaller than the inherent TFHE noise. FPT's optimized approach in which there is no need for a costly FPU or modular reduction unit allows a more lean and efficient design, coming at the cost of a one-time engineering effort to find the optimal parameters.

\begin{figure}
    \centering
    \includegraphics[width=\columnwidth]{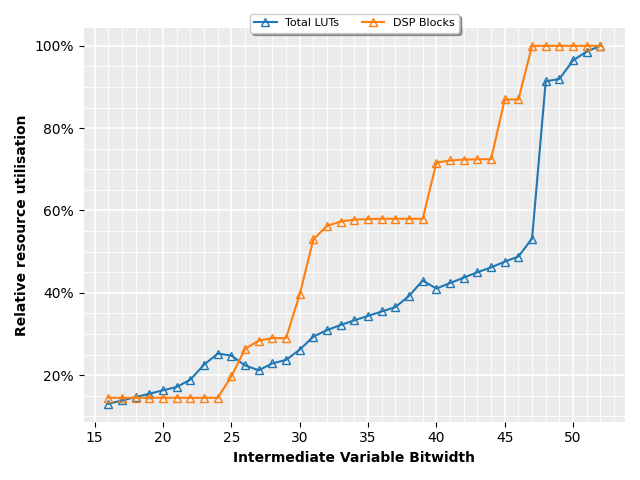}
    \caption{FPGA Relative LUT and DSP utilization of a size-256 FFT for various intermediate variable bit widths.}
    \label{fig:areavswidth}
\end{figure}

The potential effect of our fixed-point analysis on the area usage of our implementation is illustrated in \autoref{fig:areavswidth}. In this figure, we plotted the LUT and DSP usage of a size-256 FFT, as a function of the bit width of the intermediate variables. The plot gives relative numbers compared to the resource use at bit width 53 (loosely corresponding to the significand precision of IEEE 754 double-precision floating-point). As illustrated, reducing the bit width of the intermediate variables can result in a large reduction of the resource utilization, with only 20\% of the LUT and DSP usage for bit widths below 24.

Reduction of the bit width of intermediate variables relies on two parts, the location of the most significant bit, and the location of the least significant bit. We will first look at our strategy to set the MSB position of intermediate variables, and then focus on the LSB.

\subsection{Setting the MSBs}

The location of the most significant bit is important to avoid overflows. If an overflow occurs, the intermediate variable will be completely distorted and thus the result of the calculation will be unusable. Two strategies can be adopted to deal with overflows: a worst-case scenario where one can choose parameters to avoid any overflow or an average-case scenario where one allows a certain overflow with sufficiently low probability.

Avoiding any overflow comes at a significant enlargement of the parameters and thus at a significant cost, which is why we adopt the strategy to avoid overflows with a maximal overflow probability of $P_{of} = 2^{-64}$. To determine the ideal MSB position we measure the variance and then, following the central limit theorem, we assume a Gaussian distribution to calculate the overflow probability. For a given MSB position $p_{MSB}$ and standard deviation $\sigma$, the probability of overflow is:
\begin{align}
    P_{of} & = P[ |\chi| > 2^{p_{MSB}}/2 \,|\, \chi \sample \mathcal{N}(0, \sigma)] \\
           & = 1 - \erf \left( \frac{2^{p_{MSB}}}{2 \sqrt{2} \sigma} \right) .
\end{align}
Using this equation we determine the lowest $p_{MSB}$ that fulfills the maximal overflow probability of $P_{of} = 2^{-64}$ for each intermediate variable.

\subsection{Setting the LSBs}

The position of the least significant bits has an influence on the approximation noise that is introduced during the calculations. This approximation noise can be tolerated up to a certain level. More specifically, the approximation noise should be small enough so that the combination of the approximation noise and the inherent TFHE noise still leads to a correct bootstrap with high probability. We divide the total acceptable noise, for which we use theoretical noise bounds of \cite{DBLP:journals/joc/ChillottiGGI20}, into two equal parts for the approximation noise and the inherent noise, thus allowing our approximation noise to be at most half the total acceptable noise.

In our design, we focus on three main parameters: the intermediate variable widths during the forward and inverse FFT calculations, and the bit width of the coefficients of the bootstrapping key $BK$. We assume the noise introduced due to each parameter is independent (as each parameter comes from a separate block in our design), which means that the variance of the total noise is equal to the sum of the variances of each noise source ($\sigma^2_{tot} =  \sigma^2_{FFT} + \sigma^2_{IFFT} + \sigma^2_{BK}$). We then limit the noise variance due to each noise source to $1/3$th of the total noise variance.

\begin{figure}
    \centering
    \includegraphics[width=\columnwidth]{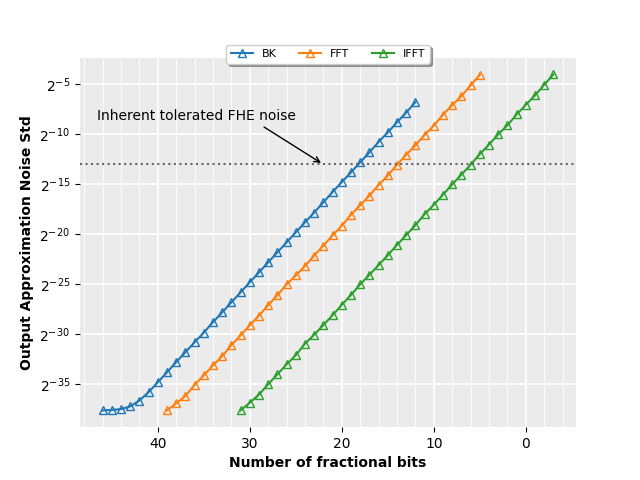}
    \caption{Output approximation noise versus the number of fractional bits for the representation of the bootstrapping key and intermediate variables during the forward and inverse FFT.}
    \label{fig:accuracy}
\end{figure}

To find optimal fixed-point parameter values, we perform a parameter sweep by setting the parameters to very high widths (in our example 53) resulting in very low noise, and then iteratively reducing one parameter until it hits the noise threshold while keeping the other parameters at high widths. The result of this experiment can be seen in \cref{fig:accuracy}, and our final fixed-point parameters are illustrated in \cref{tab:fpparams}. Note that we give the IFFT data representation before outputs are scaled by $1/N$.

\begin{table}
    \centering
    \caption{Fixed-point data representations used by intermediate variables, in the format \texttt{FixedPoint$_{\texttt{width}}$(integerBits, fractionalBits).}}
    \resizebox{\linewidth}{!}{

\begin{tabular}{l|cc}
  \toprule
  Parameter Set & I & II \\
  \midrule
  BK   &  \texttt{FixedPoint$_{\texttt{26}}$(\phantom{0}7, 19)} &  \texttt{FixedPoint$_{\texttt{27}}$(\phantom{0}8, 19)}\\
  FFT  &  \texttt{FixedPoint$_{\texttt{29}}$(15, 14)} &  \texttt{FixedPoint$_{\texttt{30}}$(18, 12)} \\
  IFFT &  \texttt{FixedPoint$_{\texttt{29}}$(23, \phantom{0}6)} &  \texttt{FixedPoint$_{\texttt{30}}$(27, \phantom{0}3)} \\
  \bottomrule
\end{tabular}
}
    \label{tab:fpparams}
\end{table}

\subsection{Related and Future Work}

One prior implementation proposing a custom hardware format for TFHE's FFTs is MATCHA~\cite{DBLP:conf/dac/0001LJ22}, who propose to use (integer) Dyadic-Value-Quantized Twiddle Factors (DVQTFs). Our fixed-point parameter analysis improves on MATCHA's in two key ways:

First, MATCHA only considers the bit width of twiddle factors, and set a uniform bit width (either 38-bit or 64-bit) that is employed throughout their external product calculations. Our analysis instead shows that different intermediate variables can profit from different fixed-point representations, allowing for an overall smaller resource utilization (\cref{fig:accuracy}, \cref{tab:fpparams}). Moreover, our analysis allows us to quantize $BK$ smaller than other parameters, limiting both on-chip $BK_i$ buffers and off-chip bandwidth.

Second, in MATCHA, instead of measuring the noise variance, the authors conduct $10^8$ tests for a parameter set to test if there are no decryption failures at the end of bootstrapping. The downside of this approach is that it becomes expensive when multiple parameters have to be set. Furthermore, this methodology does not give exact values of the failure probability, as one only has the information that no errors were found in $10^8$ tests. Our approach of measuring the approximation noise and matching it with the theoretical noise bounds provides for a more rigorous and lean design. By setting the fixed-point parameters appropriately, we can guarantee that the failure probability is arbitrarily close to the theoretical bound and floating-point implementation, and thus does not affect bootstrapping failure probability. As the server operates only on public data, moving to a carefully-tuned fixed-point representation does not affect the scheme's security.

Finally, we note that there are other intermediate variables that could be optimized, for example, the widths of the twiddle factors in the FFT calculations. We heuristically set them to the width of the intermediate variables minus 4, which gave a good balance between failure probability and cost as also explained in \cref{sec:streamingFFTs}. Interesting future work could include a full search over all possible parameters, which could result in improved fixed-point parameters over our heuristic approach.

\section{Implementation}

We implemented FPT for a AMD Alveo U280 datacenter FPGA accelerator card featuring 1.3M LUTs, 2.6M FFs, 9024 DSPs, and 41~MB of on-chip SRAM. For both parameter sets, we employ our FFT-unrolled architecture with a forward FFT streaming width of $128$ complex coefficients/clock cycle, and an IFFT streaming width of $128/l=64$ complex coefficients per clock cycle. For Parameter Set II with $N=1024$, we have also implemented the dotproduct-unrolled architecture with $(k+1)l = 4$ forward FFT kernels and $(k+2)=2$ IFFT kernels, both of uniform streaming width $32$. At this datapoint, providing iso-throughput to the FFT-unrolled architecture, we found that the dotproduct-unrolled architecture incurs 10\% more average DSP and LUT usage. Thus, we do not evaluate it further.

Our FFT-unrolled architectures feature massive throughput, completing one CMUX every $(256/128)\cdot(k+1)l=12$ clock cycles for Parameter Set I, and every $(512/128)\cdot(k+1)l=16$ clock cycles for Parameter Set II. The latency of the CMUX is larger: $144$ cycles for Parameter Set I, and $192$ cycles for Parameter Set II. In both cases, we operate at peak throughput by filling the CMUX pipeline with a batch of $b=144/12=192/16=12$ ciphertexts.

\subsection{External I/O}

The Alveo U280 includes three different host-FPGA memory interfaces: 32~GB of DDR4, 8~GB of HBM accessed through 32 Pseudo-Channels (PCs), and 24~MB of PLRAM. PBS also requires three host-side inputs: a batch of $b$ input ciphertexts $c_{in}$, the long-term bootstrapping key $BK$, and the test polynomial LUT $F$ to evaluate over the ciphertext.

For the long-term bootstrapping key, we note that it is not absolutely necessary to instantiate a ping-pong $BK_i$ buffer, as discussed in \cref{subsec:batchbootstrapping}, on our target Alveo U280 FPGA. For our parameter sets and fixed-point-trimmed $BK$ bit widths, the full $BK$ measures approximately 15~MB and fits entirely in a combination of the on-chip BRAM and URAM. Nevertheless, we instantiate a small ping-pong $BK_i$ cache as a proof-of-concept. This requires an on-chip ping-pong buffer of only $2/n$ of the full size of $BK$, allowing our architecture to remain compute-bound on architectures with less on-chip SRAM, such as smaller FPGAs or heavily memory-trimmed down ASICs. Moreover, our technique ensures that our architecture scales to new TFHE algorithms or related schemes like FHEW that increase the size of the bootstrapping key.

For our batch sizes $b$, the required $BK$ bandwidth is only tens of GB/s, which we provide by splitting the $BK$ over a limited number of HBM PCs 0-7, each providing 14~GB/s of bandwidth. The input and output ciphertext batches are small and require a negligible bandwidth, which we allocate in a single HBM PC.
Each HBM channel is served by a separate AXI master on the PL-side, which are R/W for the ciphertext and read-only for $BK$. For the test polynomial LUT $F$, we allocate an on-chip RAM that can store a configurable number of test polynomials. Each input ciphertext is tagged with an index of the LUT to apply, and correspondingly the test polynomial $F$ to select from the RAM as input to the first CMUX iteration. LUTs depend on the specific FHE program, are typically limited in number, and do not change often. For example, bootstrapped Boolean gates require only a single LUT. As such, we keep the RAM small, and we share the same HBM PC and AXI master that is used by the input and output ciphertexts.

\subsection{Xilinx Run Time Kernel}

FPT is accessible from the host as Xilinx Run Time (XRT) RTL kernel and managed through XRT API calls. FPT's CMUX pipeline features 100\% utilization during a single ciphertext batch bootstrap, and does not require complex kernel overlapping to reach peak throughput. To ensure that there are no pipeline bubbles \emph{between} the bootstrapping of different batches, we allow early pre-fetching of the next ciphertext batch into an on-chip FIFO. As such, we build FPT to support the Vitis \texttt{ap\_ctrl\_chain} kernel block level control protocol, which permits overlapping kernel executions and allows FPT to queue future ciphertext batch base HBM addresses.

\subsection{FPGA Flexibility}
FPT's microarchitecture allows explicitly tailoring the implementation to a given TFHE parameter set rather than being a one-size-fits-all approach. Instead of perceiving this as a limitation, we observe an advantage in maximizing the benefits from our target platform: FPGAs.

Because FPGAs are readily programmable, FPT is flexible to support different parameter sets and noise distributions, simply by reprogramming the FPGA. Bitstreams can be entirely pre-generated for a useful set of parameters, plaintext spaces, or even secret key distributions and associated bootstrapping methods \cite{DBLP:conf/ccs/MicciancioP21}. In its current instantiation, we provide bitstreams for parameter sets I and II, with more variations expected as future work. Reprogramming the FPGA is fast, and expected to happen only infrequently within a given application. As such, our approach allows us to heavily optimize the accelerator for each parameter set while remaining flexible to support a large variety of TFHE applications.

\subsection{Fixed-point Streaming Design in Chisel}

While the outer host-FPGA communication logic of FPT is implemented in SystemVerilog, we use Chisel \cite{DBLP:conf/dac/BachrachVRLWAWA12} -- an open-source HDL embedded in Scala -- to construct the inner streaming CMUX kernel. Like SystemVerilog, Chisel is a full-fledged HDL with direct constructs to describe synthesizable combinational and sequential logic and not a High-Level Synthesis (HLS) language. Our motivation to select Chisel over SystemVerilog for the CMUX, is that it makes the full capabilities of the Scala language available to describe circuit generators. We make heavy use of object-oriented and functional programming tools to describe our CMUX streaming architecture for a configurable streaming width, and in both realizations shown in \cref{fig:FPT}. Moreover, Chisel has a rich type system that is further supported by external libraries. In FPT, the existing \texttt{DspComplex[FixedPoint]} is our main hardware datatype that we use within our architecture. Building on existing \texttt{FixedPoint} test infrastructure that we extended for FPT, our experiments in \cref{sec:fixedpointparams} are directly run on the Chisel-generated Verilog rather than an intermediate fixed-point software model.

\section{Evaluation and Comparison}

\begin{table}
    \centering
    \caption{FPT's FPGA Resource Utilization Breakdown for Parameter Sets I and II. The high utilization of DSP blocks make them the limiting resources.}
    \begin{tabular}{l|S[table-format=4.0]|S[table-format=4.0]|S[table-format=4.0]|S[table-format=4.0]}
  \toprule
                              & \thdrl{LUT (K)}   &   \thdrl{FF (K)} &  \thdrl{DSP} &  \thdr{BRAM} \\
  \midrule
  \multirow{2}{*}{FPT - \textit{Param. Set I}}
                              &       {40\% av.}  &       {35\% av.} &   {61\% av.} &   {25\% av.} \\
                              &               526 &              916 &         5494 &          505 \\
  \quad     CMUX              &               384 &              707 &         5494 &          310 \\
  \quad\quad MAC (384\ttimes) &                97 &              114 &         2304 &          310 \\ 
  \quad\quad FFT$_{256,128}$  &               159 &              366 &         2126 &            0 \\
  \quad\quad IFFT$_{256,64}$  &                97 &              192 &         1064 &            0 \\
  \arrayrulecolor{gray!50}\midrule
  \multirow{2}{*}{FPT - \textit{Param. Set II}}
                              &       {46\% av.}  &       {39\% av.} &   {66\% av.} &   {20\% av.} \\
                              &               595 &             1024 &         5980 &          412 \\
  \quad     CMUX              &               458 &              827 &         5980 &          215 \\
  \quad\quad MAC (256\ttimes) &                66 &               79 &         1536 &          215 \\
  \quad\quad FFT$_{512,128}$  &               222 &              449 &         2958 &            0 \\
  \quad\quad IFFT$_{512,64}$  &               130 &              255 &         1486 &            0 \\
  \arrayrulecolor{black}\bottomrule
\end{tabular}
    \label{tab:utilization}
\end{table}

\subsection{Resource Utilization}
FPT is implemented using AMD Vivado 2022.2 and packaged as XRT kernel using Vitis 2022.2, targeting a clock frequency of \SI{200}{\mega\hertz}. \cref{tab:utilization} presents a resource utilization breakdown of FPT, for both Parameter Sets I and II. In both cases, DSP blocks are the main limiting resource that prevents increasing to the next available streaming width, with up to 66\% of available DSP blocks utilized by FPT. Note that whereas \cref{fig:FPT} presented our ping-pong $BK$ buffer as a monolithic memory block, it is physically split into many smaller memory blocks placed inside the MAC units that consume them.

\begin{table*}
    \centering
    \caption{Comparison of TFHE PBS on a Variety of Platforms}
    \begin{tabular}{l|c|lrrrr|S[table-format=4.0]|S[table-format=1.2]|S[table-format=1.2]}
  \toprule
                                  & Parameter Set & \thdr{Platform} & \thdr{LUT} & \thdr{FFs} & \thdr{DSP} & \thdrl{BRAM}                & {Clock (MHz)} & {Latency(ms)} & {TP (PBS/ms)} \\
  \midrule
  \multirow{2}{*}{\textbf{FPT}}   & I             & \multirow{2}{*}{FPGA}
                                                                  & 526 K      & 916 K      & 5494       & 17.5 Mb                     &         200 &        0.66{$^\dagger$} & 28.4 \\
                                  & II            &               & 595 K      & 1024 K     & 5980       & 14.5 Mb                     &         200 &        0.74{$^\dagger$} & 25.0 \\ 
  \midrule 
                                  &               & \multirow{2}{*}{FPGA}
                                                                  & 842 K      & 662 K      & 7202       & 338 Mb                      &         180 &        3.76 & 3.5 \\
  \multirow{-2}{*}{YKP \cite{DBLP:conf/hpec/YeKP22}} 
                                  & \multirow{-2}{*}{II} 
                                                  &               & 442 K      & 342 K      & 6910       & 409 Mb                      &         180 &        1.88 & 2.7 \\
  \midrule
  \midrule 
  
  \multirow{1}{*}{MATCHA \cite{DBLP:conf/dac/0001LJ22}} 
                                  & \multirow{1}{*}{II} 
                                                  & ASIC          & \multicolumn{4}{l|}{\SI{36.96}{\milli\meter\squared} \SI{16}{\nano\meter} PTM}                    &        2000 &         0.2 & 10.0 \\
  \midrule
  \multirow{2}{*}{CONCRETE \cite{chillotti2020concrete}}
                                  & I             & \multirow{2}{*}{CPU}
                                                                  & \multicolumn{4}{l|}{\multirow{2}{*}{Intel Xeon Silver 4208}} &         2100 &         33.0 & 0.03 \\
                                  & II            &               &                                                        & & & &         2100 &         32.0 & 0.03 \\
  \midrule
  \multirow{2}{*}{TFHE-rs \cite{tfhers}}
                                  & I             & \multirow{2}{*}{CPU}
                                                                  & \multicolumn{4}{l|}{\multirow{2}{*}{Intel Xeon Silver 4208}} &         2100 &         9.25 & 0.11 \\
                                  & II            &               &                                                        & & & &         2100 &         9.45 & 0.11 \\
  \midrule
  cuFHE \cite{DBLP:conf/balkancryptsec/DaiS15,cuFHE} 
                                  & II            & GPU           & \multicolumn{4}{l|}{NVIDIA GeForce RTX 3090}                  &        1700 &          9.34 & 9.6 \\
  \bottomrule
  \multicolumn{10}{p{0.95\textwidth}}{}\\[-2mm]
  \multicolumn{10}{p{0.95\textwidth}}{
    \textit{$^\dagger$ The latency is identical to complete a single or a batch of $b=12$ PBS.}}  
\end{tabular}

    \label{tab:performance}
\end{table*}

\subsection{PBS Benchmarks}
\label{sec:pbsbench}

\cref{tab:performance} compares FPT quantitatively with a number of prior TFHE baselines. For our CPU baseline, we benchmark single-core PBS on an Intel Xeon Silver 4208 CPU at \SI{2.1}{\giga\hertz} for both CONCRETE \cite{chillotti2020concrete}, and its recent update TFHE-rs \cite{tfhers} employing a heavily-optimized FFT library. A recent ASIC baseline is provided by MATCHA \cite{DBLP:conf/dac/0001LJ22}. MATCHA \textit{emulates} a \SI{36.96}{\milli\meter\squared} ASIC in a \SI{16}{\nano\meter} PTM process technology. As an FPGA baseline, we include a recent architecture of Ye et al. \cite{DBLP:conf/hpec/YeKP22} (referred to as YKP after the authors' initials), which was developed concurrently with our work and significantly improves the prior baseline of Gener et al. \cite{gener2021fpga}. Lastly, we also include in our comparison YKP's benchmarks of cuFHE \cite{DBLP:conf/balkancryptsec/DaiS15,cuFHE}, a GPU-based implementation benchmarked on an NVIDIA GeForce RTX 3090 GPU at \SI{1.7}{\giga\hertz}.

FPT is optimized for maximal throughput. \cref{tab:performance} illustrates FPT's massive throughput, enabled through its streaming architecture: 937\ttimes higher than CONCRETE, 7.1\ttimes higher than YKP, and 2.5\ttimes higher than MATCHA or cuFHE.

In the current version of FPT, latency optimization was not prioritized. Many TFHE homomorphic applications feature hundreds to thousands of parallel ciphertexts. In such settings, the total time to bootstrap all ciphertexts --governed by the throughput-- is much more important than the latency to bootstrap a single ciphertext. FPT's PBS latency is determined by the PCIe and AXI latencies of communicating input and output ciphertexts\footnote{These latencies are amortized by overlapping kernel executions. In \cref{tab:performance}, they account for the difference between Latency and $(b=12)$/TP.}, as well as its CMUX pipeline depth. In this work, we kept the CMUX pipeline depth large, fitting $b=12$ ciphertexts and enabling small off-chip bandwidth through our batched bootstrapping technique.

Lower-latency implementations of FPT can opt to decrease the CMUX pipeline depth, requiring either more off-chip memory bandwidth to load $BK$ or caching the full $BK$ on-chip. Although these lower-latency practices are not included in the current implementation, FPT's latency is still competitive with MATCHA.

Finally, FPT is estimated at 99W total on-chip power (FPGA and HBM), offering a similar TP/W as MATCHA (\SI{40}{\watt}) and significantly more than cuFHE ($>$\SI{200}{\watt}) or YKP (\SI{50}{\watt}).

\begin{table}[b]
    \caption{
        The Performance Comparison of FPT over the Game of Life Demo. The FPT accelerated board updates much quicker than the native software counterpart.}
    \begin{tabular}{l|c|S[table-format=1.2]|S[table-format=1.2]|S[table-format=1.2]}
    \toprule
                                & Key     & \multicolumn{1}{>{\centering\arraybackslash}p{1.9cm}|}{Board}   & {Bootstrap}       & {Key Switch     } \\
                                & Switch  & {Update (s)}                                                    & {2816\ttimes (s)} & {2816\ttimes (s)} \\
    \midrule 
    \multirow{2}{*}{TFHE-rs}    & with    & 29.1                                                            & 26.7              & 2.3               \\
                                & w/out   & 47.3                                                            & 47.3              &                   \\
    \midrule
                                & with    &  2.51 {\hspace*{2.5mm}\textit{12\ttimes}}                       & 0.11              & 2.3               \\
    \multirow{-2}{*}{FPT}       & w/out   &  0.3  {\hspace*{2.5mm}\textit{97\ttimes}}                       & 0.26              &                   \\
    \bottomrule
\end{tabular}

    \label{table:gol}
\end{table}

\subsection{Conway's Game of Life}
\label{sec:gol}

To evaluate FPT in a more comprehensive application; we benchmarked an encrypted simulation of Conway's Game of Life \cite{optalysys_gol}. The Game of Life is a simulation of a two-dimensional board, where each board cell is either in the \emph{alive} or \emph{dead} state. At each time step, a cell's state transitions depending on its neighbors' state. Many different types of patterns can appear on the board, including still-lifes, oscillators, or patterns that travel across the board.

In running Game of Life over TFHE, the server receives an encrypted initial board configuration with encrypted cell states. Update rules are translated into Boolean equations, which are calculated by the server using encrypted gate arithmetic \cite{optalysys_gol}. Updating a single cell state requires exactly 44 encrypted gate computations, disregarding the cheaper NOT gates that do not include a bootstrap. As a whole, the encrypted Game of Life consists of a mix of homomorphic AND, XOR, OR, and NOT gates.
These operations and their parallel computation should help estimating FPT's performance on a variety of applications. In addition, this is an application which demonstrates the performance improvements live: the FPGA-accelerated board updates visually appear much quicker than the software counterpart.

We made the artifacts of our demo available\footnote{Artifacts available at: \url{https://github.com/KULeuven-COSIC/fpt-demo}} to be run on AWS F1 (FPGA Instances), together with a screenrecording of its execution. Notice that, this artifact is only a demo version that prefers a reduced parameter set to have the software TFHE-rs board updates in acceptable time.

We benchmarked homomorphic Game of Life inside the most recent TFHE-rs software library for an $8 \times 8$ board. For this board size, a single time step update requires exactly $64\times44=2816$ programmable bootstraps. Inside TFHE-rs, we integrated FPT as a backend for Parameter Set I. Using simple library function calls, the CPU can offload the bootstrap calculations to the FPGA. As one caveat, our current first integration into TFHE-rs does not allow overlapping kernel executions. Because of this limitation, FPT's throughput drops from $28.4$ PBS/ms to $12/0.66=18.2$ PBS/ms, still 165\ttimes higher than TFHE-rs.

The results of our experiment are summarized in \cref{table:gol}. In the TFHE-rs software version, a board update takes \SI{29.1}{\second}, of which \SI{26.7}{\second} are spent in bootstrapping, \SI{2.3}{\second} seconds in key-switching, and merely \SI{0.1}{\second} in calculating other operations. A straightforward integration of FPT offers a moderately limited speedup, quickly facing Amdahl's law: a board update is reduced to \SI{2.5}{\second} (10.7\ttimes), of which now only \SI{0.11}{\second} are spent in bootstrapping, but \SI{2.3}{\second} in key-switching. In this setting, because the CPU spends most of the time in key-switching instead of providing ciphertexts to FPT, the FPGA stays idle 96\% of the time.

In addition to considering the key-switching as the next operation to accelerate, we can algorithmically exploit FPT's high bootstrapping throughput. Specifically, key-switching can entirely be eliminated by increasing the TLWE input dimension $n=586$ to match the TLWE output dimension $kN=1024$\footnote{Increasing $n$ increases the security.}. However, this technique makes bootstrapping roughly $1024/586=1.75$\ttimes as expensive. In software, this change is undesired: board update times increase to \SI{47.3}{\second}, 99.9\% of which is bootstrapping. In contrast, FPT's high bootstrapping shines in this setting: board updates are reduced to \SI{0.3}{\second} only. Here, FPT achieves a total 158\ttimes speedup over TFHE-rs and is active about 60\% of the time.

In summary, in its current integration into TFHE-rs, FPT can accelerate an application containing a mix of encrypted gate calculations by a factor of $29.1 / 0.3 \simeq 100$\ttimes.

More generally, these results should readily extend from Game of Life to other TFHE applications that are bottlenecked by PBS computations. For example, in \cite{DBLP:conf/cscml/ChillottiJP21}, the authors show how encrypted deep neural networks can make heavy use of PBS. The resulting encrypted networks incur a slowdown factor of $\sim$\SI{1}{\mega\nothing} over their unencrypted counterparts. Utilizing FPT, this slowdown factor could be reduced to $\sim$\SI{10}{\kilo\nothing} instead.

Finally, we note that a 100\ttimes acceleration is less than what FPT promises, with 258\ttimes higher bootstrapping throughput than TFHE-rs. Maximizing these throughput gains of FPT requires devoting increased engineering effort on a given application. Nevertheless, to our knowledge, FPT is one of the first accelerators to be integrated and demonstrated end-to-end within a popular FHE software library.

\begin{table*}[t]
    \centering
    \caption{Performance and Cost Comparison of FPT against instances on AWS with/without GPU-acceleration}
    \begin{center}
  \begin{tabular}{ p{4.5cm}|S[table-format=5.0]S[table-format=5.0]|S[table-format=5.0]S[table-format=5.0]|S[table-format=5.0]S[table-format=5.0] }
    \toprule
                                                & \multicolumn{2}{c|}{FPT}
                                                                                      & \multicolumn{2}{c|}{GPU Acceleration$^*$}
                                                                                                                              & \multicolumn{2}{c}{Without Acceleration} \\
                                                & \multicolumn{1}{c}{Local Setup}      
                                                                  & \multicolumn{1}{c|}{AWS F1}            
                                                                                      & \multicolumn{1}{c}{AWS P4     }
                                                                                                          & \multicolumn{1}{c|}{AWS P3           } 
                                                                                                                              & \multicolumn{1}{c}{AWS EC2 M6}
                                                                                                                                                 & \multicolumn{1}{c}{AWS EC2 C4} \\
                                                & \multicolumn{1}{c}{Alveo U280}      
                                                                  & \multicolumn{1}{c|}{Custom}            
                                                                                      & \multicolumn{1}{c}{NVIDIA A100}
                                                                                                          & \multicolumn{1}{c|}{NVIDIA Tesla V100} 
                                                                                                                              & \multicolumn{1}{c}{Intel Xeon}
                                                                                                                                                 & \multicolumn{1}{c}{Intel Xeon} \\
    \midrule  
    FPGA~/~GPU~/~CPU Count                      & 1     & 1      & 8      & 1      & 2     & 36    \\
    Million Operations per Day$^\star$              & 2200  & 2200   & 3975   & 352    & 16    & 294   \\
    Monthly \$ Cost with Discounts$^\dagger$    & 567   & 1060   & 19140  & 1930   & 62    & 1060  \\
    Million Operations per \$                   & 117   & 63     & 6      & 5      & 8     & 8     \\
    \midrule
    \multicolumn{7}{l}{\textit{The number of instances required to reach FPT's performance and their cost:}}                                                         \\
    Number of Instances                         & 1     & 1      & 1      & 7      & 135   & 8     \\
    Monthy \$ Cost                              & 567   & 1060   & 19140  & 1930   & 8425  & 8460  \\
    Factor of Cost                              &       & 2      & 34     & 24     & 15    & 15    \\
    \bottomrule
    \multicolumn{7}{p{0.95\textwidth}}{}\\[-2mm]
    \multicolumn{7}{p{0.95\textwidth}}{\textit{$*$ Performance is estimated by a linear interpolation of \cite{cuFHE, DBLP:conf/hpec/YeKP22}'s benchmarks over the target GPUs' CUDA core count and clock frequency.}} \\
    \multicolumn{7}{p{0.95\textwidth}}{\textit{$\star$ Assuming n times performance improvement from n FPGAs~/~GPUs~/~CPUs.}} \\
    \multicolumn{7}{p{0.95\textwidth}}{\textit{$\dagger$ AWS prices of July 19th, 2023. A 20\% reduction is applied for renting the instances for three years with in-advance payments. For the local version, setup cost is distributed among an expected 5 years lifespan, and electricity/internet cost is added based on tariffs in Belgium.}} \\
  \end{tabular}
  \end{center}

    \label{tab:cost}
\end{table*}

\subsection{Cost Comparison}

To more comprehensively evaluate FPT against CPU or GPU-based execution, FPT's performance advantages are elevated with computation cost benefits summarized in \cref{tab:cost}. The table compares the operating cost of FPT on a local FPGA server against various Amazon Web Services (AWS) instances that consists of FPGA, GPU or only CPUs. For fairness, CPU and GPU-based execution are given advances, such as maximum benefit from parallel threads, AWS offered price discounts, etc. \cref{tab:cost} indicates a monthly \$ cost of 15\ttimes to reach FPT's performance with the other execution methods.

We note that not all TFHE applications require the peak bootstrapping throughput of FPT. In a time-varying setting where FPT is used only 50\% of the time on average, CPU and GPU-based execution can instead take advantage by utilizing fewer or more time-varying compute instances. However, even in this setting, there still is a $\gtrapprox$10\ttimes cost advantage.

Last but not least, it is also worth considering that FPGAs do not require being used in PCIe accelerator card form. Rather, they can be built using custom compute servers, tailored to custom data and compute distribution networks, as exemplified by Microsoft Catapult \cite{DBLP:conf/iiswc/Chiou17}.

\subsection{Related Work}

Qualitatively, FPT makes different design choices than either YKP or MATCHA. MATCHA is built after the classical CPU approach to FHE accelerators. It includes a set of TGGSW clusters with external product cores that operate from a register file. As one result, MATCHA is bottlenecked by data movement and cache memory access conflicts.

YKP is an HLS-based architecture that redefines TFHE to use NTT, breaking compatibility with existing TFHE libraries like CONCRETE, and disabling the fixed-point optimizations of FPT. At the architectural level, YKP includes some concepts also employed by FPT. Similar to FPT, they include a pipelined implementation of the CMUX that processes multiple ciphertext instances. However, unlike FPT which builds a single streaming CMUX PE with large and configurable streaming width, YKP implements and instantiates multiple smaller CMUX PEs with inferior TP/A. Each CMUX pipeline instance in YKP includes an SRAM that stores a coefficient $BK_i$. However, unlike FPT where these SRAMs are loaded by off-chip memory in ping-pong fashion, YKP loads coefficients from DRAM only after a full coefficient has been consumed. This makes the number of CMUX PEs they instantiate limited by the off-chip memory bandwidth, whereas FPT's design choices make it entirely compute-bound.

Both MATCHA and YKP focus on an algorithmic technique called \emph{bootstrapping key unrolling}. This technique unrolls $m$ iterations of the Blind Rotation loop (\cref{alg:PBS}, \cref{line:pbsblindrotate}), requiring an (exponentially) more expensive CMUX equation and larger $BK$, but reducing the total number of iterations from $n$ to $n/m$. Bootstrapping key unrolling was not considered for FPT, but is a promising future technique to evaluate. Moreover, since FPT is not bound by off-chip memory bandwidth, more aggressive key unrolling could be feasible.

For completeness, we note that both MATCHA and YKP include key-switching as an operation of PBS. Key-switching includes coefficient-wise multiplication of a TLWE ciphertext with a key-switching key. We opted not to include key-switching in FPT, because different FHE programs may choose to key-switch either before or after PBS \cite{DBLP:journals/joc/ChillottiGGI20}, and key-switching can be circumvented for an accelerator as described in \cref{sec:gol}. Nevertheless, key-switching is an operation with much lower throughput requirements than the CMUX \cite{DBLP:conf/hpec/YeKP22}. In FPT, key-switching of the output ciphertext can be supported without throughput penalty by instantiating a few integer multipliers on the AXI write-back path. Finally, we note that neither MATCHA nor YKP integrate with an FHE software library, or provide benchmarks for a complete application.

\section{Conclusion}

In this paper, we introduced FPT, an accelerator for the Torus Fully Homomorphic Encryption (TFHE) scheme. In contrast to previous FHE architectures, our design follows a streaming approach with high throughput and low control overhead. Owing to a batched design and balanced streaming architecture, our accelerator is the first FHE bootstrapping implementation that is compute-bound and not memory-bound, with small data caches and a 100\% utilization of the arithmetic units. Instead of using an NTT or floating-point FFT, FPT achieves a significant throughput increase by utilizing up to 80\% area-reduced fixed-point FFTs with compact and optimized variable representations. In the end, FPT achieves a TFHE bootstrapping throughput of 28.4 bootstrappings per millisecond, which is 258\ttimes-947\ttimes faster than CPU implementations, 7.1\ttimes faster than a concurrent FPGA implementation, and 2.5\ttimes faster than state-of-the-art ASIC and GPU designs.

\begin{acks}
  This work was supported in part by CyberSecurity Research Flanders with reference number VR20192203, the Research Council KU Leuven (C16/15/058), the Horizon 2020 ERC Advanced Grant (101020005 Belfort), and the AMD University Program through the donation of a AMD Alveo U280 datacenter accelerator card.

  Michiel Van Beirendonck is funded by FWO as Strategic Basic (SB) PhD fellow (project number 1SD5621N). Jan-Pieter D'Anvers is funded by FWO (Research Foundation - Flanders) as junior post-doctoral fellow (contract number 133185).

  Finally, the authors would like to thank Wouter Legiest for experimenting with a variety of FFT generator tools.
\end{acks}

\FloatBarrier

\bibliographystyle{ACM-Reference-Format}
\bibliography{dblp,googlescholar}

\end{document}